\documentclass[twoside,twocolumn,american,preprintnumbers,nofootinbib,aps,pra,10pt]{revtex4-1}

\pdfoutput=1

\usepackage{graphicx}
\usepackage{float}
\usepackage{slashed}
\usepackage[font=small, labelfont=bf]{caption}
\usepackage[labelformat=simple]{subcaption}

\usepackage{amsmath, amsthm, amssymb, amsfonts}
\usepackage{multirow}
\usepackage{hyperref}

\def\beq{\begin{align}}
\def\eeq{\end{align}}
\newcommand{\bi}{\begin{itemize}}
\newcommand{\ei}{\end{itemize}}
\newcommand{\ben}{\begin{enumerate}}
\newcommand{\een}{\end{enumerate}}
\newcommand{\be}{\begin{equation}}
\newcommand{\ee}{\end{equation}}
\newcommand{\bea}{\begin{eqnarray}}
\newcommand{\eea}{\end{eqnarray}}

\begin{document}

\title{New accelerating solutions in late-time cosmology}

\author{Michele Cicoli$^{1,2}$, Giuseppe Dibitetto$^3$, Francisco G. Pedro$^{1,2}$}
\affiliation{$^{1}$Dipartimento di Fisica e Astronomia, Universit\`a di Bologna, via Irnerio 46, 40126 Bologna, Italy}
\affiliation{$^{2}$INFN, Sezione di Bologna, viale Berti Pichat 6/2, 40127 Bologna, Italy}
\affiliation{$^{3}$Department of Physics, University of Oviedo, Avda.  Federico Garcia Lorca s/n, 33007 Oviedo, Spain}


\begin{abstract}
Dark energy models can be seen as dynamical systems. In this paper we show that multi-field models with a curved field space give rise to new critical points and we analyse their stability. These are new accelerating solutions in late-time cosmology which exist even for steep potentials. This opens up the possibility to realise quintessence models even when quantum corrections spoil the flatness of the underlying potential. These non-linear sigma models arise naturally in supergravity and string models where their multi-field dynamics can help to avoid swampland bounds.
\end{abstract}

\maketitle

\section{Introduction}

Right before the turn of the millenium the ground-breaking detection of dark energy completely changed our view of our universe. A combination of different observations \cite{Riess:1998cb, Tegmark:2003ud, Jaffe:2000tx}, gave birth to the so-called $\Lambda$CDM model, according to which roughly 70\% of our universe is currently driving a phase of accelerated expansion.

This late-time behaviour of our universe is described by cosmological models which can be cast in the form of dynamical systems. This allows to study their fixed points and their stability via both analytical and numerical methods \cite{Tsujikawa:2013fta}. So far several solutions have been found mainly for single-field models. These include matter domination, kinetic domination, scaling solutions and accelerating solutions where the universe is dominated by the potential energy of a scalar field. More general solutions can be found by including multiple scalar fields mainly focusing on a flat field space (see \cite{Bahamonde:2017ize} and references therein).\footnote{See \cite{vandeBruck:2009gp} for a preliminary discussion of the curved field space case.} In this case no qualitatively new critical points emerge but the regions of stability can be enlarged. 

In analogy with existing inflationary solutions \cite{Brown:2017osf,Christodoulidis:2019jsx}, in this paper we shall instead consider multi-field models of late-time cosmology where the field space is curved. We shall show that these non-linear sigma models give rise to qualitatively new fixed points where accelerated expansion is possible even for steep potentials. This opens up the possibility of realising viable dark energy models in general cases where quantum corrections would otherwise spoil the flatness of standard quintessence solutions. 

In particular, we will consider models with two scalar fields where $\phi_1$ has an exponential potential, $\phi_2$ is a flat direction and the model features a field-dependent kinetic coupling between these two fields.\footnote{The inflationary dynamics of very similar models has already been studied in \cite{Cicoli:2018ccr,Cicoli:2019ulk}.} Starting from arbitrary initial conditions corresponding to matter domination, the system evolves following a spiral trajectory where a non-zero velocity along $\phi_2$ can be seen as an extra time-dependent effective contribution to the potential of $\phi_1$. This effect is crucial to generate a late-time accelerating solution which can reproduce the observed equation of state parameter and energy density of dark energy even if the exponential potential for $\phi_1$ is steep. 

Notice that non-linear sigma models of this kind emerge naturally from string compactifications \cite{Cicoli:2018ccr,Cicoli:2019ulk,Krippendorf:2018tei} and represent a general class of dark energy constructions which seem promising to evade recently conjectured quantum gravity bounds. In fact, the existence of consistent de Sitter (dS) vacua in quantum gravity is currently under fervid debate \cite{Brennan:2017rbf,Danielsson:2018ztv,Ooguri:2018wrx,Cicoli:2018kdo}. This resulted in the so-called `dS swampland conjecture' \cite{Obied:2018sgi}, stating the existence of $\mathcal{O}(1)$ bounds on the slope of effective potentials which is in tension with standard realisations of quintessence models with flat potentials. Our new accelerating solutions would help to avoid these bounds by relying on a multi-field dynamics in a curved field space.\footnote{See \cite{Achucarro:2018vey}for similar considerations in the context of inflation.}

\section{Quintessence from field space curvature}

\subsection{A two-field model}

In this paper we consider the late universe dynamics of a two-field scalar sector of the form:
\be
\frac{\mathcal{L}}{\sqrt{-g}} =\frac12 \,\gamma_{ij}\partial_\mu \phi^i \partial^\mu \phi^j- V(\phi) \ , 
\label{eq:model}
\ee
where the field space metric can be written as:
\be
\gamma_{ij}=\begin{pmatrix}
1 & 0 \\
0 & f^2(\phi_1) 
\end{pmatrix}
\label{gammaij}
\ee
and where the potential has a flat direction along $\phi_2$, i.e. $V(\phi)=V(\phi_1)$. This class of metrics occurs often in string and supergravity constructions where the two fields are part of the same super-multiplet. The specific form of the function $f$ will depend on the geometric origin of the scalars. In what follows we will show that this system can give rise to a quasi-dS phase in the late universe where the expansion of spacetime is driven by the non-trivial dynamics of the massless scalar. 

The two-field system is described by the coupled Klein-Gordon equations:
\be
\left\{
\begin{array}{ll}
\ddot{\phi}_1+3H\dot{\phi}_1-f\,f_1 \dot{\phi}_2^2+V_1=0\\[5pt]
\ddot{\phi}_2+3H\dot{\phi}_2+2\frac{f_1}{f}\dot{\phi}_2\dot{\phi}_1=0
\end{array}.\right.
\ee
In order to obtain exact analytical results we will study the dynamics of the system when the scalar potential is of exponential form:
\be
V=V_0 \, e^{-k_2\phi_1}\,,
\ee
and the field-space curvature is constant and negative:
\be
R=-|R| = -2\,\frac{f_{11}}{f}=-2 k_1^2\ ,
\label{eq:eqf}
\ee
which implies a kinetic coupling of the form:
\be
f(\phi_1)=A_+ \,e^{k_1 \phi_1}+A_-\,e^{-k_1\phi_1}\,.
\label{fform}
\ee 
For concreteness we will take $A_+=0$ and set $A_-=1$. In order to study the dynamics of the system of coupled scalars in the late universe we also add a barotropic fluid with pressure $p_\gamma=(-1+\gamma)\rho_\gamma$ that evolves according to the continuity equation $\dot{\rho}_\gamma=-3 H \gamma \rho_\gamma$.
We assume $0<\gamma<2$ and will often take the barotropic fluid to be pressureless dust and set $\gamma=1$. The first Friedmann equation becomes:
\be
H^2=\frac{1}{3 M_P^2}\left(\frac{\dot{\phi}_1^2}{2}+\frac{f^2}{2}\dot{\phi}_2^2+V+\rho_\gamma\right) \ .
\label{eq:Fried}
\ee
Following \cite{Copeland:1997et} we define the dimensionless variables:
\be
x_1\equiv \frac{\dot{\phi}_1}{\sqrt{6} H M_p}\,,\quad x_2\equiv \frac{f \dot{\phi}_2}{\sqrt{6} H M_p}\,,\quad y_1\equiv \frac{\sqrt{V}}{\sqrt{3} H M_p}\,,
\ee
which allow us to write the dynamics of the system as an autonomous system (with $' \equiv d/d \ln a$):
\be
x_1' = h_1\,,\qquad  x_2'=h_2\,,\qquad y_1'=h_3\,,
\label{AutSyst}
\ee
where:
\bea
\frac{h_1}{x_1} &=& 3 \left(x_1^2+ x_2^2-1 \right) + \sqrt{\frac32} \left(-2 k_1 x_2^2 + k_2 y_1^2\right) x_1 \nonumber \\
&-& \frac32 \gamma \left( x_1^2 + x_2^2 + y_1^2-1 \right) \nonumber \\
\frac{h_2}{x_2} &=& 3\left(x_1^2+x_2^2-1\right) +\sqrt{6} k_1 x_1 -\frac32 \gamma  \left(x_1^2+x_2^2+y_1^2-1\right) \nonumber \\
\frac{h_3}{y_1} &=& -\sqrt{\frac32} k_2 x_1-\frac32 \gamma  \left(x_1^2+x_2^2+y_1^2-1\right)+3 \left(x_1^2+x_2^2\right).
\nonumber
\eea

Given that we are assuming a flat universe, eq. \eqref{eq:Fried} can be written as $ \Omega_\gamma=1-(x_1^2+x_2^2+y_1^2)$ with $x_1, x_2 \in [-1,1]$ and $y_1\in [0,1]$. The physical parameter space, when written in terms of these variables, is half of a 3-disk (or a ball) of unit radius centred at the origin. 

The equation of state for the scalar sector is a crucial variable for late universe physics and can be written as:
\be
\omega_\phi=\frac{p_\phi}{\rho_\phi}=\frac{x_1^2+x_2^2-y_1^2}{x_1^2+x_2^2+y_1^2}\,,
\ee
while the energy density in the scalar sector is: 
\be
\Omega_\phi=x_1^2+x_2^2+y_1^2 \,.
\ee
Current observations \cite{Aghanim:2018eyx} point to  $\omega_\phi\simeq-1$ and $\Omega_\phi \simeq 0.7$ today, with exact values and corresponding errors varying depending on the datasets used. 

\begin{figure}[!h]
  \centering
    \includegraphics[width= 0.40\textwidth]{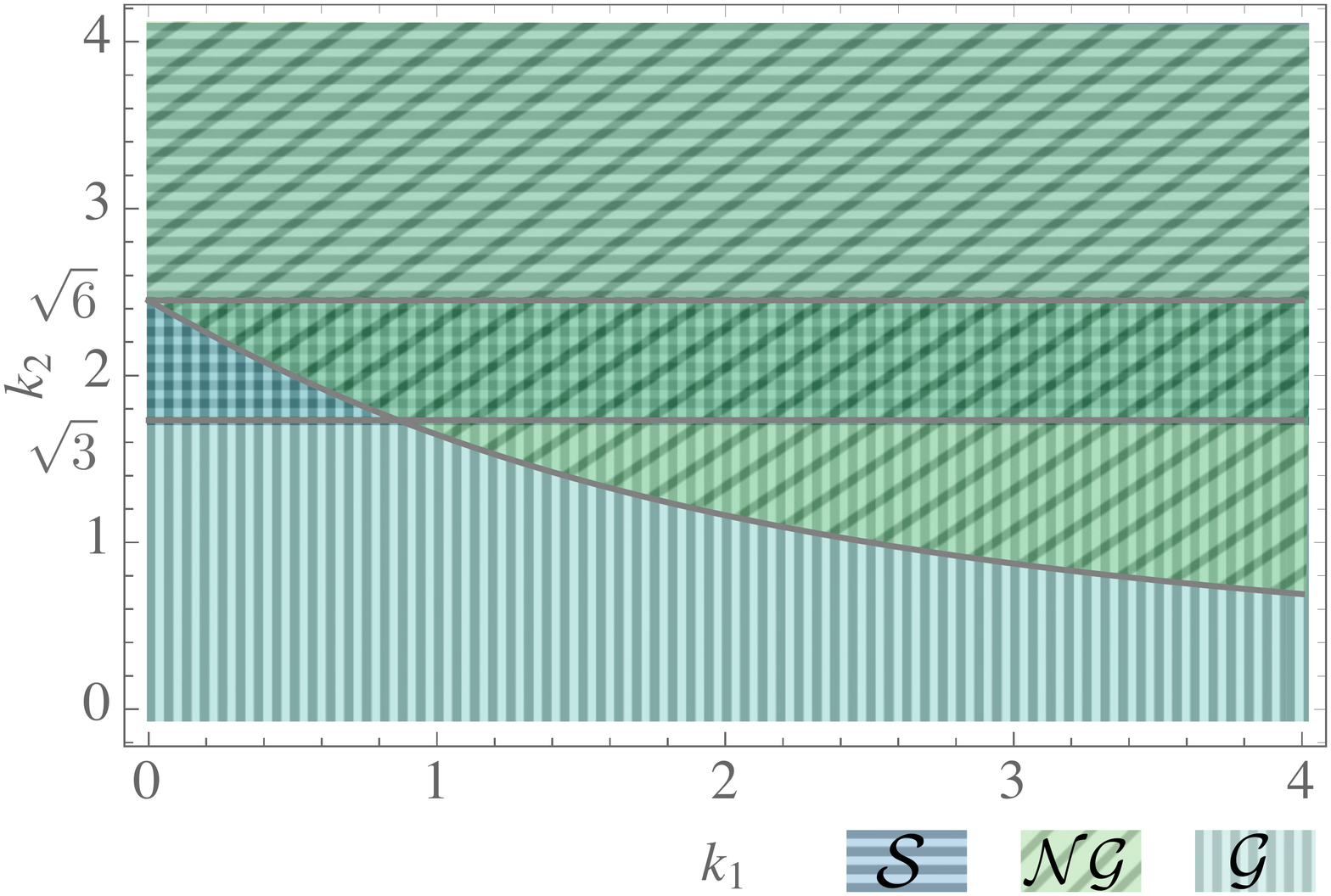}
  \caption{Existence domains for the $\mathcal{G}$, $\mathcal{NG}$ and  $\mathcal{S}$ fixed points in a matter background with $\gamma=1$.}
\label{fig:Existence}
\end{figure}

The critical points of the system (\ref{AutSyst}) correspond to $ x_1'= x_2'= y_1'=0$.  We find six critical points which we list in Tab. \ref{tab:fp} and whose non-trivial existence domains we illustrate in Fig. \ref{fig:Existence}.

\begin{table*}[htp]
\begin{center}
\begin{tabular}{c||c|c|c||c|c|c}
&$x_1$&$x_2$&$y_1$ & $\Omega_\phi$ & $\omega_\phi$ & existence \\
\hline
\hline
$\mathcal{K}_+$&$1$ & $0$  & $0$ & 1   & 1  & all $ k_1, k_2, \gamma$ \\
$\mathcal{K}_-$&$-1$ & $0$  & $0$ & 1   & 1   & all $k_1, k_2, \gamma$ \\
$\mathcal{F}$&$0$ & $0$  & $0$ & 0   & undefined & all $k_1, k_2, \gamma$   \\
$\mathcal{S}$&$\frac{\sqrt{3/2}\gamma}{k_2}$ & $0$ & $\frac{\sqrt{3/2\gamma(2-\gamma)}}{k_2}$ & $\frac{3 \gamma}{k_2^2}$ & $\gamma-1$ & $0<\gamma<2 \wedge  k_2^2\ge 3 \gamma$\\
$\mathcal{G}$&$ \frac{k_2}{\sqrt{6}}$ & $0$  & $\sqrt{1-\frac{k_2^2}{6}}$ & $1$   & $-1+\frac{k_2^2}{3} $& $k_2<\sqrt{6}$\\
$\mathcal{NG}$&$\frac{\sqrt{6}}{(2k_1+k_2)}$ & $\frac{\pm \sqrt{k_2^2+2k_2 k_1-6}}{2 k_1 + k_2}$ & $\sqrt{\frac{2 k_1}{2k_1+k_2}}$ & $1$ & $\frac{k_2-2k_1}{k_2+2 k_1}$ &$k_2 \ge  \sqrt{6 + k_1^2}-k_1$\\
\hline
\end{tabular}
\end{center}
\caption{Fixed points of the system with a barotropic fluid and two scalar fields.}
\label{tab:fp}
\end{table*}

Two critical points, $\mathcal{K_+}$ and $\mathcal{K}_-$, correspond to kinetic domination, while $\mathcal{F}$ to fluid domination. For the critical point $\mathcal{G}$ (where $\mathcal{G}$ stands for `geodesic' since the system evolves along a geodesic trajectory in field space) the universe is dominated by the scalar potential energy, while for the scaling solution, $\mathcal{S}$, the scalar sector tracks the fluid energy density, with $\Omega_\phi/\Omega_\gamma$ constant. In these critical points the flat direction $\phi_2$ plays no r\^ole as is evident from the fact that all have $x_2=0$. These critical points are well known in single-field models \cite{Copeland:1997et}. The $\mathcal{NG}$ critical point (where $\mathcal{NG}$ now stays for `non-geodesic' since the evolution is along a non-geodesic trajectory in field space) is the novel feature of this model arising from the presence of the flat direction $\phi_2$. We see that in this critical point $\phi_2$ is dragged by the kinetic energy of $\phi_1$. The interesting aspect of this new fixed point is that $\omega_\phi\simeq -1$ can be achieved even if the scalar potential sourcing $H$ is too steep. The key ingredient is pushing the system into a strong curvature regime with $k_1\gg k_2$. This could be an interesting mechanism to build quintessence models that are compatible with swampland constraints on flat potentials from string theory.

\begin{figure}[!h]
  \centering
    \includegraphics[width= 0.40 \textwidth]{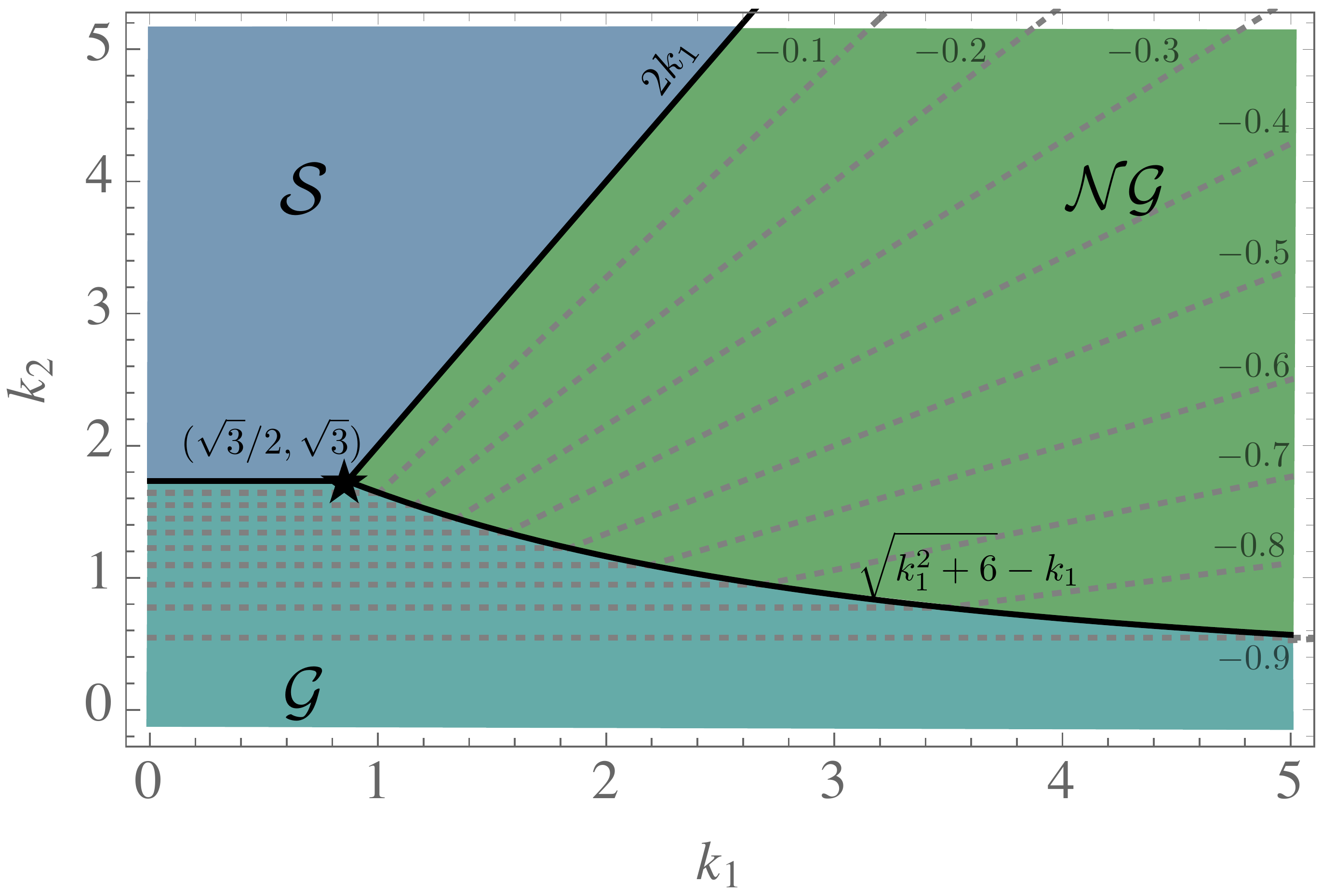}
  \caption{Stability diagram for $\gamma=1$. Dotted lines denote the value of the equation of state parameter $\omega_\phi$.}
\label{fig:phaseDiag}
\end{figure}

Besides establishing the existence of fixed points for the system (\ref{AutSyst}), one can determine their stability by studying perturbations around them. The time evolution of the perturbations is determined by the eigenvalues of the stability matrix $M_{ij}=\partial h_i/\partial x_j$ with $i,j=1,2,3$ and $x_3\equiv y_1$. In the appendix we list the eigenvalues of $M_{ij}$ and the stability conditions that can be derived from them, which we depict in Fig. \ref{fig:phaseDiag} for $\gamma=1$. It is evident that for any value of $k_2$ a sufficiently large $k_1$ can give a stable $\mathcal{NG}$ fixed point with $\omega_\phi\simeq -1$. In other words one can have accelerated expansion regardless of the steepness of the potential by going into the regime where the field-space curvature is large.

\subsection{Viable dark energy solutions}
\label{sec:trans}

Having established the existence and stability properties of the fixed points, we now try to employ them to describe the late-time universe. Given that observations point to $\omega_\phi\simeq -1$ and $\Omega_\phi\simeq 0.7$ we automatically see that none of the fixed points above can be viable: only $\mathcal{S}$ allows for $\Omega_\phi<1$ but it does not give rise to accelerated expansion for the most natural situation $\gamma=1$. One is therefore forced to model the current state of the universe as a transient phase that will eventually lead the system to one of the aforementioned fixed points.

\begin{figure}[!h]
  \centering
    \includegraphics[width= 0.4\textwidth]{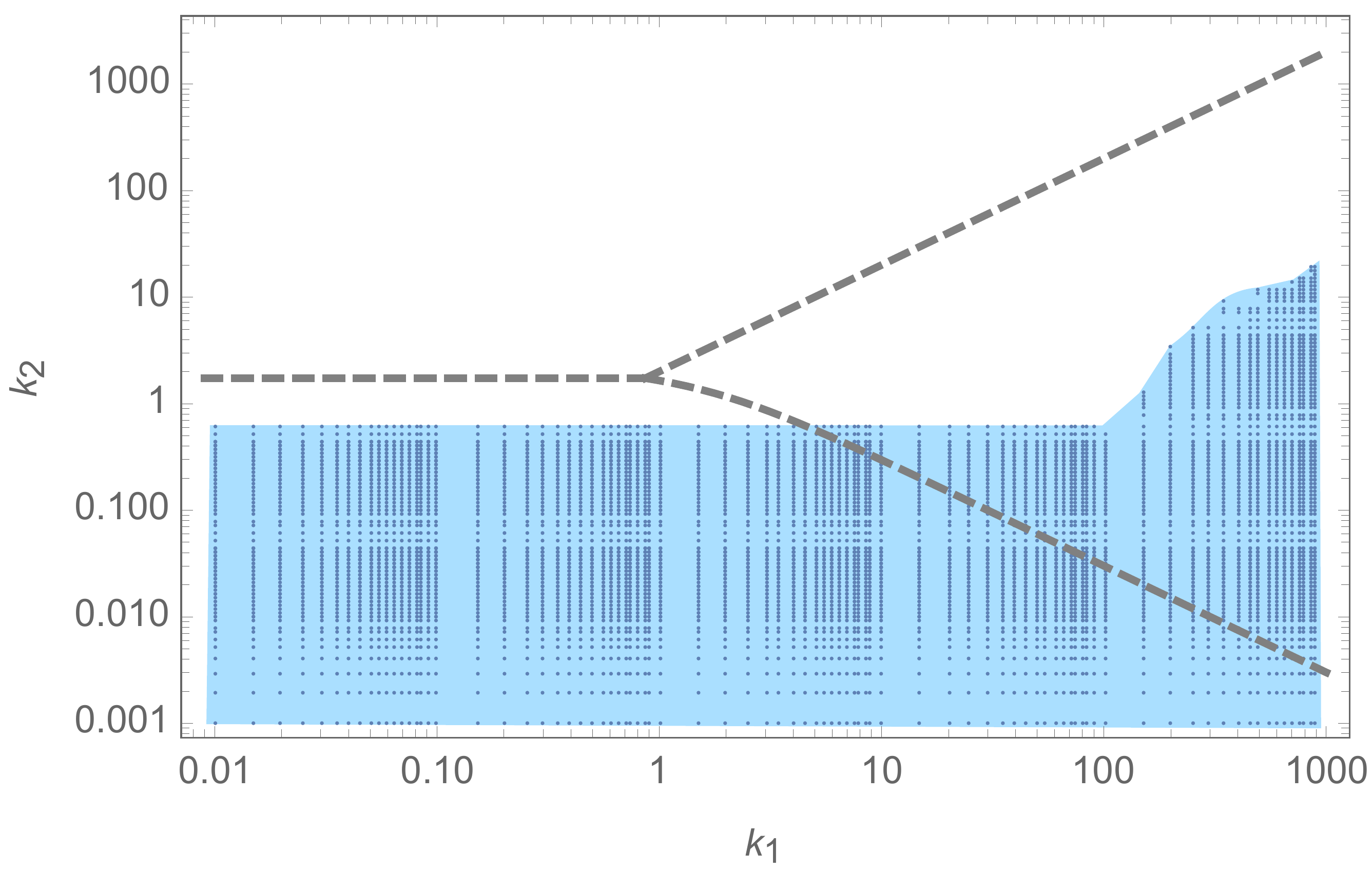}
  \caption{Viable region in the $(k_1,k_2)$ plane that for matter dominated initial conditions $(x_1,x_2,y_1)= (10^{-4},10^{-4},10^{-4})$ feature a point with $\Omega_\phi=0.7$ and $\omega_\phi$ in agreement with the bounds from the Pantheon sample \cite{Scolnic:2017caz} as presented in \cite{Agrawal:2018own, Akrami:2018ylq}. Dashed lines represent the boundaries between the stability domain of each fixed point.}
\label{fig:viableTrans}
\end{figure}

With that in mind we perform a scan of the $(k_1,k_2)$ plane looking for transients through $\omega_\phi\simeq -1$ and $\Omega_\phi\simeq 0.7$. We choose matter ($\gamma=1$) dominated initial conditions $(x_1,x_2,y_1)\sim (0,0,0)$. For each point in parameter space we look for $\Omega_\phi\simeq 0.7$ and then assess if $\omega_\phi$ is in agreement with the bounds derived in \cite{Scolnic:2017caz} for redshifts below unity, using the parametrisation \cite{Chevallier:2000qy,Linder:2002et} $\omega_\phi=\omega_0+\omega_a \frac{z}{1+z}$. The results of this numerical analysis are presented in Fig. \ref{fig:viableTrans}. Notice that viable transients can be attained in a large part of parameter space, not only when the potential is flat, $k_2\le 0.6$, but also for $k_2 \ge 1$, provided $k_1$ is sufficiently large. Note that for flat potentials we reproduce the results of \cite{Agrawal:2018own}, where the universe is today in a transient evolving towards $\mathcal{G}$. In this regime the two-field dynamics changes only the future endpoint and has little bearing on the present evolution. If the field-space curvature is sufficiently large to allow for $\mathcal{NG}$ to exist and be stable, the system upon reaching the unstable $\mathcal{G}$ fixed point evolves along $\Omega_\phi=1$ trajectories towards $\mathcal{NG}$. Conversely for small curvature only $\mathcal{G}$ exists and is stable, erasing any signs of two-field dynamics.
 
\begin{figure}[!h]
  \centering
      \includegraphics[width= 0.4\textwidth]{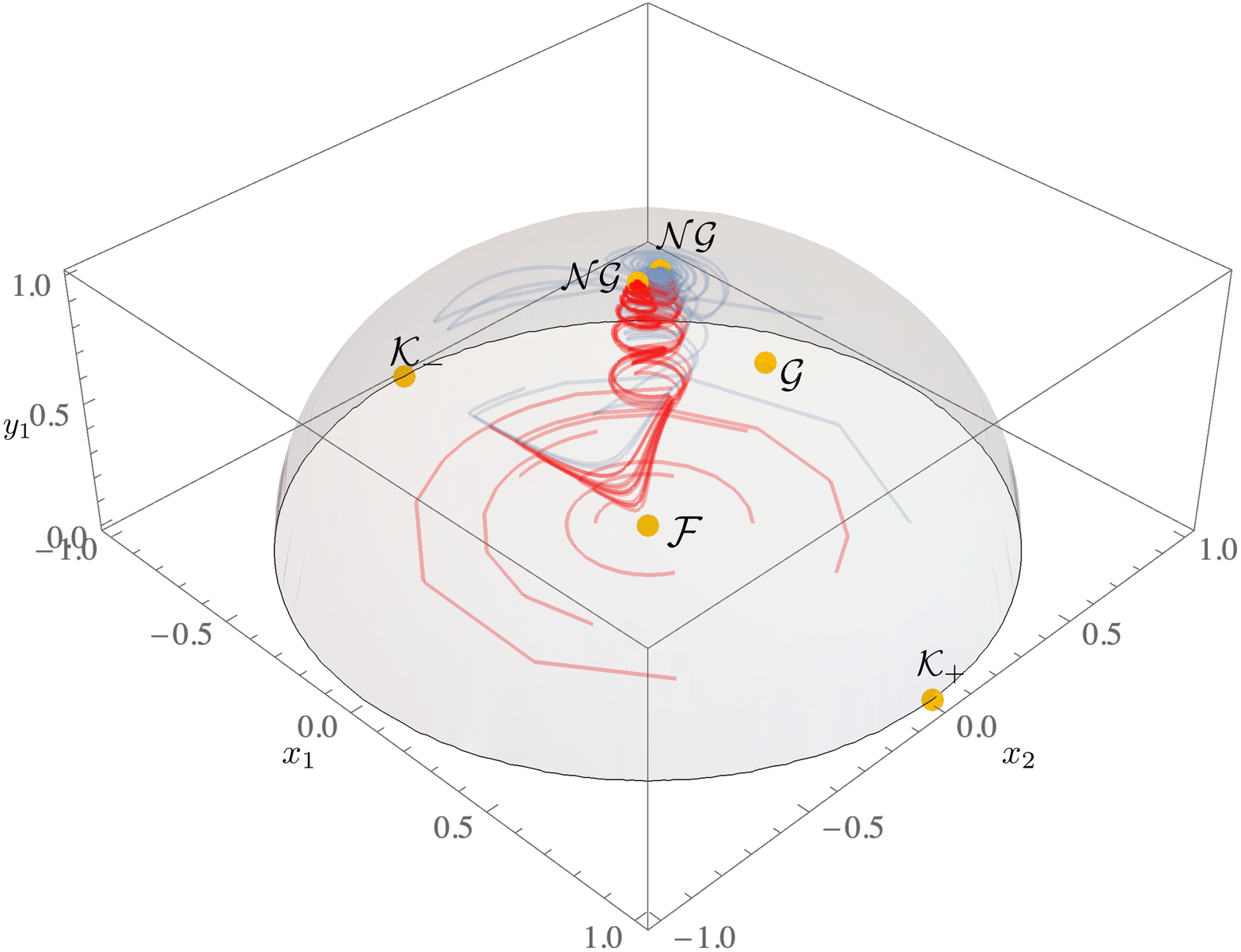}
    \includegraphics[width= 0.4\textwidth]{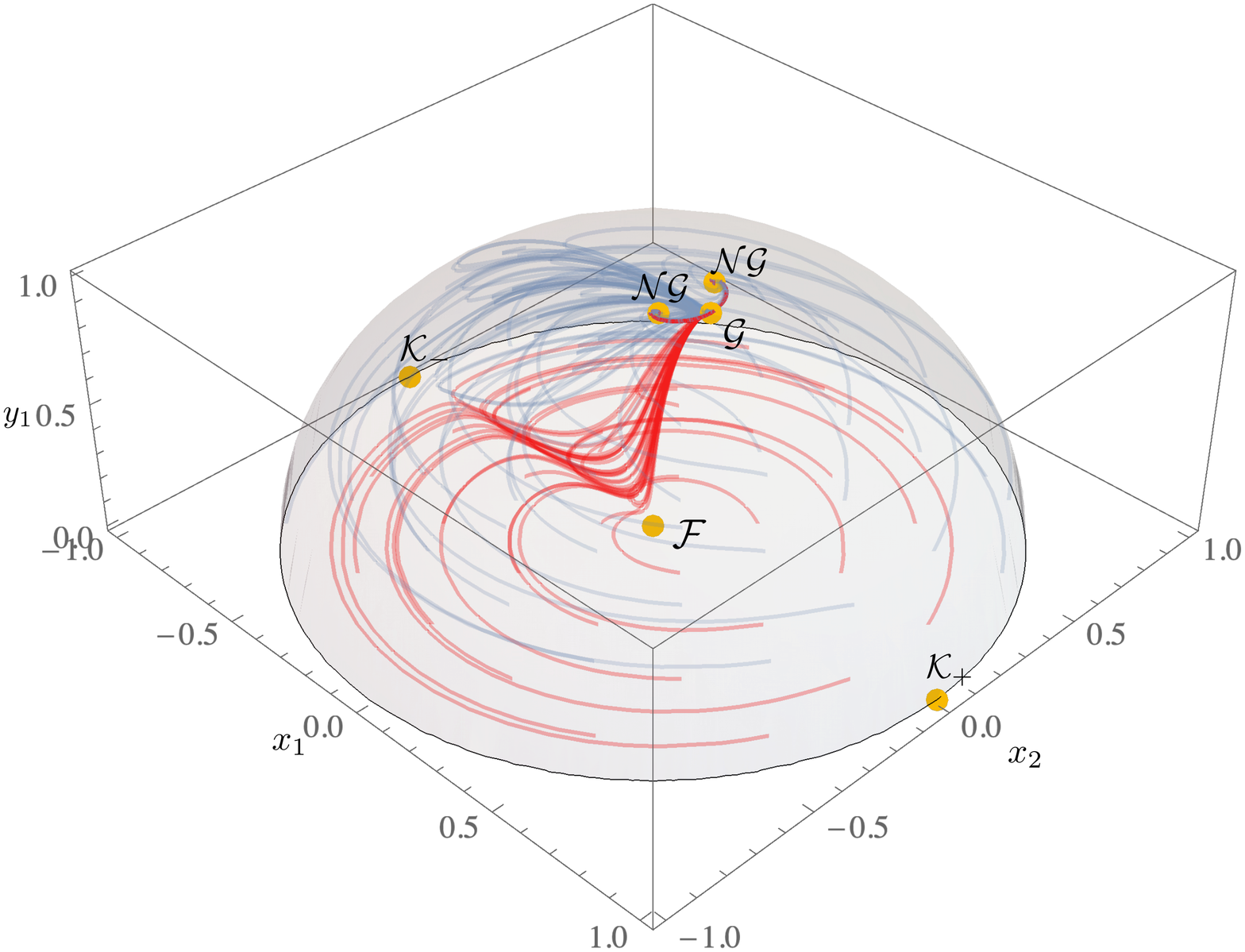}
  \caption{Observationally viable phase space trajectories. Top: $k_1=300$, $k_2=1$; bottom: $k_1=10$, $k_2=1/2$.}
\label{fig:traj}
\end{figure}

The novel feature of the two-field setup is the existence of a dark energy regime with $\omega_\phi \simeq -1$ and $\Omega_\phi\simeq 0.7$ even when the scalar potential is steep. Viable models in this regime feature a stable $\mathcal{NG}$ fixed point together with unstable $\mathcal{G}$ and/or $\mathcal{S}$ fixed points (depending on the steepness of the potential). Observationally viable transients are obtained as the system evolves from matter domination directly towards $\mathcal{NG}$ in the vicinity of the $(x_1,x_2)=(0,0)$ axis.
In Fig. \ref{fig:traj} we display the time evolution of the system for a range of initial conditions corresponding to matter domination. For $k_1=10$ and $k_2=1/2$ the system evolves towards $\mathcal{G}$, which in this case is a saddle, before settling into the pair of $\mathcal{NG}$ attractors. On the other hand, for $k_1=300$ and $k_2=1$, the distance between $\mathcal{G}$ and the two $\mathcal{NG}$ fixed points increases and the system simply evolves towards the non-geodesic solution. Red lines represent trajectories which can yield $\Omega_\phi\simeq 0.7$ and $\omega_\phi\simeq -1$, while blue trajectories do not satisfy present observational bounds.

\section{Summary}

In this paper we proposed a new mechanism to generate a period of late-time accelerated expansion. Contrary to existing mechanisms, it does not rely on a flat potential, but rather on a curved field space in the presence of a flat direction. We performed a stability analysis of the minimal two-field setup, showing that it reproduces all ingredients of known single-field setups and it features an extra fixed point where the equation of state is not uniquely determined by the flatness of the scalar potential. Viable quintessence models can be built using this new fixed point for any given scalar potential by tuning the field space curvature large, thereby attaining a transient through $\omega_\phi\simeq-1$ and $\Omega_\phi\simeq 0.7$.

This can open the way to build new quintessence models in agreement with observations for scalar potentials whose flatness is lifted by quantum corrections. Let us also stress that these non-linear sigma models arise rather generically in string constructions and seem promising frameworks to build dark energy models in agreement with constraints from the swampland program. It would be very interesting to find explicit string models with all moduli stabilised except for a runaway direction which could play the r\^ole of our $\phi_1$ field. $\phi_2$ might instead be its axionic partner that would be naturally ultra-light since its mass is protected at perturbative level by a shift symmetry. We leave this investigation for future work. 

We finally mention that quintessence models are notoriously plagued by problems with fifth forces which would be shared also by our model. In fact, if $\phi_2$ is a pseudoscalar, as expected from string models, it would not cause any problem. However $\phi_1$, if it couples to ordinary matter with Planckian strength, would need to be as heavy as at least $\mathcal{O}(1)$ meV, which would require $k_2\gtrsim 10^{30}$. To avoid such a huge value of $k_2$ one has to suppress the couplings of $\phi_1$, for example by geometrical sequestering in the extra dimensions as in \cite{Cicoli:2012tz, Acharya:2018deu}.

\section*{Appendix}

In this appendix we display the eigenvalues of the stability matrix $M_{ij}$ (denoted by $\mu$), from which we derive the stability conditions for each fixed point.
\begin{itemize}
\item $\mathcal{K}_\pm$: For $k_1,k_2>0$ and $0<\gamma<2$, all fixed points are either saddles or unstable since:
\be
\mu=\left \{\ \pm \sqrt{6} k_1\ ,\  \frac{1}{2} (6 \mp \sqrt{6} k_2)\ , \ 3 (2 - \gamma)\  \right \} \nonumber
\ee

\item $\mathcal{F}$: This fixed point is a saddle for $0<\gamma<2$ as:
\be
\mu=\frac{3}{2}\left \{ \ -2 + \gamma \ ,  -2 + \gamma \ , \gamma\ \right \} \nonumber
\ee

\item $\mathcal{S}$: 
For the phenomenologically interesting $\gamma=1$ case one has:
\be
\mu=\left\{-\frac{3}{2}+\frac{3 k_1}{k_2},-\frac{3}{4}\left(1\pm \frac{ \sqrt{24-7 k_2^2}}{k_2}\right)\right\}\ , \nonumber
\ee
which leads to the following stability region:
\be
k_2> 2 k_1 \quad\text{and}\quad k_2>\sqrt{3}\,. 
\ee

\item $\mathcal{G}$: For $\gamma=1$ and $k_1>0$ this fixed point is stable in the region:
\be
0<k_2<\sqrt{3}\quad\text{and}\quad 0<k_2<\sqrt{k_1^2+6}-k_1 \nonumber
\ee
since:
\be
\mu=\left\{\frac{1}{2} \left(k_2^2-6\right),k_1 k_2+\frac{k_2^2}{2}-3,k_2^2-3 \gamma \right\} \nonumber
\ee

\item $\mathcal{NG}$: In this case the exact eigenvalues cannot be computed analytically, however we may note that for this fixed point to provide a viable description of dark energy, $\omega_\phi\sim-1$ and therefore $\omega_\phi+1=\frac{2 k_2}{2 k_1+k_2}\ll1$. To zeroth order in $\omega_\phi+1$ one finds: 
\be
\mu\simeq \left \{ \ -3, \frac{1}{2} \left (-3 \pm \sqrt{3} \sqrt{27 - 8 k_1 k_2 - 4 k_2^2}\right) \right \} \nonumber
\ee
and so stability can be achieved for $k_2 \ge -k_1 +\sqrt{6 + k_1^2}$, which matches the existence condition in Tab. \ref{tab:fp}. To get a complete picture of the stability of this fixed point one has to use numerical methods to compute the eigenvalues of $M_{ij}$ for generic $k_1$ and $k_2$, finding that stability can be achieved for: 
\be  
\sqrt{6 + k_1^2}-k_1\le k_2\le 2 k_1\ . 
\label{eq:stabNG}
\ee
This numerical result can be confirmed analytically by checking that the stability matrix develops a vanishing eigenvalue along the borders of the stability region defined in eq. \eqref{eq:stabNG}.
\end{itemize}


\begin{thebibliography}{10}  

\bibitem{Riess:1998cb}
  A.~G.~Riess {\it et al.} [Supernova Search Team],
  Astron.\ J.\  {\bf 116} (1998) 1009
  doi:10.1086/300499

\bibitem{Tegmark:2003ud}
  M.~Tegmark {\it et al.} [SDSS Collaboration],
  Phys.\ Rev.\ D {\bf 69} (2004) 103501
  doi:10.1103/PhysRevD.69.103501

\bibitem{Jaffe:2000tx}
  A.~H.~Jaffe {\it et al.} [Boomerang Collaboration],
  Phys.\ Rev.\ Lett.\  {\bf 86} (2001) 3475
  doi:10.1103/PhysRevLett.86.3475

\bibitem{Tsujikawa:2013fta}
  S.~Tsujikawa,
  Class.\ Quant.\ Grav.\  {\bf 30} (2013) 214003
  doi:10.1088/0264-9381/30/21/214003

\bibitem{Bahamonde:2017ize}
  S.~Bahamonde, C.~G.~Böhmer, S.~Carloni, E.~J.~Copeland, W.~Fang and N.~Tamanini,
  Phys.\ Rept.\  {\bf 775-777} (2018) 1
  doi:10.1016/j.physrep.2018.09.001

\bibitem{vandeBruck:2009gp}
  C.~van de Bruck and J.~M.~Weller,
  Phys.\ Rev.\ D {\bf 80} (2009) 123014
  doi:10.1103/PhysRevD.80.123014

\bibitem{Brown:2017osf}
  A.~R.~Brown,
  Phys.\ Rev.\ Lett.\  {\bf 121} (2018) no.25,  251601
  doi:10.1103/PhysRevLett.121.251601

\bibitem{Christodoulidis:2019jsx}
  P.~Christodoulidis, D.~Roest and E.~I.~Sfakianakis,
  JCAP {\bf 1912} (2019) no.12,  059
  doi:10.1088/1475-7516/2019/12/059

\bibitem{Cicoli:2018ccr}
  M.~Cicoli, V.~Guidetti, F.~G.~Pedro and G.~P.~Vacca,
  JCAP {\bf 1812} (2018) 037
  doi:10.1088/1475-7516/2018/12/037

\bibitem{Cicoli:2019ulk}
  M.~Cicoli, V.~Guidetti and F.~G.~Pedro,
  JCAP {\bf 1905} (2019) 046
  doi:10.1088/1475-7516/2019/05/046

\bibitem{Krippendorf:2018tei}
  S.~Krippendorf, F.~Muia and F.~Quevedo,
  JHEP {\bf 1808} (2018) 070
  doi:10.1007/JHEP08(2018)070

\bibitem{Brennan:2017rbf}
  T.~D.~Brennan, F.~Carta and C.~Vafa,
  PoS TASI {\bf 2017} (2017) 015
  doi:10.22323/1.305.0015

\bibitem{Danielsson:2018ztv}
  U.~H.~Danielsson and T.~Van Riet,
  Int.\ J.\ Mod.\ Phys.\ D {\bf 27} (2018) no.12,  1830007
  doi:10.1142/S0218271818300070

\bibitem{Ooguri:2018wrx}
  H.~Ooguri, E.~Palti, G.~Shiu and C.~Vafa,
  Phys.\ Lett.\ B {\bf 788} (2019) 180
  doi:10.1016/j.physletb.2018.11.018

\bibitem{Cicoli:2018kdo}
  M.~Cicoli, S.~De Alwis, A.~Maharana, F.~Muia and F.~Quevedo,
  Fortsch.\ Phys.\  {\bf 67} (2019) no.1-2,  1800079
  doi:10.1002/prop.201800079

\bibitem{Obied:2018sgi}
  G.~Obied, H.~Ooguri, L.~Spodyneiko and C.~Vafa,
  arXiv:1806.08362 [hep-th].

\bibitem{Achucarro:2018vey}
  A.~Achúcarro and G.~A.~Palma,
  JCAP {\bf 1902} (2019) 041
  doi:10.1088/1475-7516/2019/02/041

\bibitem{Copeland:1997et}
  E.~J.~Copeland, A.~R.~Liddle and D.~Wands,
  Phys.\ Rev.\ D {\bf 57} (1998) 4686
  doi:10.1103/PhysRevD.57.4686

\bibitem{Aghanim:2018eyx}
  N.~Aghanim {\it et al.} [Planck Collaboration],
  arXiv:1807.06209 [astro-ph.CO].

\bibitem{Scolnic:2017caz}
  D.~M.~Scolnic {\it et al.},
  Astrophys.\ J.\  {\bf 859} (2018) no.2,  101
  doi:10.3847/1538-4357/aab9bb

\bibitem{Chevallier:2000qy}
  M.~Chevallier and D.~Polarski,
  Int.\ J.\ Mod.\ Phys.\ D {\bf 10} (2001) 213
  doi:10.1142/S0218271801000822

\bibitem{Linder:2002et}
  E.~V.~Linder,
  Phys.\ Rev.\ Lett.\  {\bf 90} (2003) 091301
  doi:10.1103/PhysRevLett.90.091301

\bibitem{Agrawal:2018own}
  P.~Agrawal, G.~Obied, P.~J.~Steinhardt and C.~Vafa,
  Phys.\ Lett.\ B {\bf 784} (2018) 271
  doi:10.1016/j.physletb.2018.07.040

\bibitem{Akrami:2018ylq}
  Y.~Akrami, R.~Kallosh, A.~Linde and V.~Vardanyan,
  Fortsch.\ Phys.\  {\bf 67} (2019) no.1-2,  1800075
  doi:10.1002/prop.201800075
  
\bibitem{Cicoli:2012tz}
  M.~Cicoli, F.~G.~Pedro and G.~Tasinato,
  JCAP {\bf 1207} (2012) 044
  doi:10.1088/1475-7516/2012/07/044
		
\bibitem{Acharya:2018deu}
  B.~S.~Acharya, A.~Maharana and F.~Muia,
  JHEP {\bf 1903} (2019) 048
  doi:10.1007/JHEP03(2019)048
	
  
\end{thebibliography}
\end{document}